\documentclass[11pt]{article}

\usepackage{myArticle, subfig}
\graphicspath{{../figures/}}
\usetikzlibrary{decorations.pathreplacing,arrows, arrows.meta}

\newcommand{\x}{\mathbf{x}}

\title{A Cardinality Minimization Approach to Security-Constrained Economic Dispatch}
\author[1]{David~Troxell\thanks{dtroxell@smu.edu}}
\author[1]{Miju Ahn\thanks{mijua@smu.edu}}
\author[1]{Harsha Gangammanavar\thanks{harsha@smu.edu}}
\date{First submission: April, 2021\\Current version: November 9, 2021}

\affil[1]{Department of Operations Research and Engineering Management, Southern Methodist University, Dallas TX}

\begin{document}\thispagestyle{empty}
\maketitle

\begin{abstract}
We present a threshold-based cardinality minimization formulation to model the security-constrained economic dispatch problem. The model aims to minimize the operating cost of the system while simultaneously reducing the number of lines operating in emergency operating zones during contingency events. The model allows the system operator to monitor the duration for which lines operate in emergency zones and ensure that they are within the acceptable reliability standards determined by the system operators. We develop a continuous difference-of-convex approximation of the cardinality minimization problem and a solution method to solve the problem. Our numerical experiments demonstrate that the cardinality minimization approach reduces the overall system operating cost as well as avoids prolonged periods of high electricity prices during contingency events.
\end{abstract}

\newcommand{\f}{\mathbf{f}}
\newcommand{\change}[1]{{\color{black}#1}}

\section{Introduction} \label{sect:intro}
The independent system operators (ISOs) and regional transmission organizations (RTOs) oversee the non-discriminatory access to transmission assets, operate the transmission system independently, and foster competitive generation among wholesale market participants. The ISO/RTOs use bid-based markets arranged hierarchically over multiple timescales. The lowest timescale of the market scheduling process is the real-time Security-Constrained Economic Dispatch (SCED). SCED is used to determine generation across the power system to satisfy the electricity demand at the least cost while meeting the system reliability requirements.

SCED aims to balance the intrinsically competing goals of system efficiency and reliability. While system efficiency is realized by fully utilizing the available transmission capacity, system reliability requires conserving transmission capacity to handle contingency scenarios. Maintaining this balance in real-time poses a significant challenge to the system operators in normal operating conditions, let alone system operations in extreme weather events such as the Texas winter storm in February 2021. The large-scale integration of intermittent renewable resources such as wind and solar also exasperates system operations.

SCED is modeled as a single-period or multiperiod optimization model. Since it was first proposed in \cite{alsac1974optimal}, the SCED optimization models have evolved significantly. See \cite{Chowdhury1990} for a review of early and \cite{capitanescu2011state,frank2012optimal} for a review of more recent works. This evolution is driven by the desire to include more detailed system operations (e.g., inclusion of AC optimal power flow \cite{wu2019security}), reliability requirements (e.g., transient stability \cite{abhyankar2017solution}), and market considerations (e.g., flexible ramping \cite{wang2014flexible}), as well as computational considerations. 

SCED can be performed either in preventive or corrective forms \cite{pappu2013optimization}. The preventive approach aims to determine a base-case dispatch solution that can withstand contingency scenarios without any adjustments. On the other hand, the corrective approach allows the base-case dispatch solution to deviate cost-effectively. While the preventive approaches are more secure, they tend to be overly conservative. Therefore, approaches that aim to determine a base-case dispatch solution that minimizes deviation under presumed contingency are preferred. In these approaches, it is also desirable to treat transmission as a flexible asset. In this regard, SCED base-case dispatch solutions can be determined by considering corrective transmission switching or corrective rescheduling. 

Corrective transmission switching allows a transmission element to be switched out of service shortly after a contingency occurs to avoid post-contingency violations. Transmission switching has many benefits such as improved reliability \cite{li2016real, shao2005corrective}, congestion management \cite{khodaei2010transmission}, and ease the incorporation of renewable resources \cite{korad2013robust}. In the context of SCED, corrective transmission switching has been studied recently in \cite{li2019enhanced}.

Corrective rescheduling using mathematical optimization was first proposed in \cite{monticelli1987security}. The underlying assumption of this approach is that the operational limit violations (e.g., thermal limits of transmission lines) can be endured for limited periods.  For example, a given power line can have a normal rating (denoting the most desirable thermal rating), a long-term emergency (LTE) rating, and a short-term emergency (STE) rating \cite{Douglass2019}. Prior corrective rescheduling works have focused on the contingency filtering that aims to reduce the number of contingency scenarios to be considered for determining the base-case dispatch solution \cite{capitanescu2008new, strbac1998method}. More recently, a multistage contingency response model was proposed in \cite{liu2014computational}. Since corrective rescheduling is undertaken as a reactive measure, the models proposed in these works implicitly consider the knowledge of when the contingency occurs. Moreover, these models impose a bound on the deviation from the base-case at a given period and ignore the duration of time for which the limit violations occur. For instance, New England ISO requires the line flow to return to less than the LTE rating within $15$ minutes \cite{caiso2020establishing}. Incorporating these reliability requirements in SCED results in a non-convex optimization problem (even when we consider linear/direct current approximations or convex relaxations of the power flow). In light of these, the main contributions of our work are as follows:
\begin{enumerate}
	\item \emph{SCED as cardinality minimization problem.} We introduce a SCED formulation with an objective function that includes the operational costs and a reliability term that penalizes operating the transmission lines in emergency zones. The model is intended for corrective rescheduling following a contingency event. The model accommodates differential penalties for operating in tiered emergency operating zones and includes strict constraints on the duration of time a particular line operates in an emergency zone. The resulting model is a threshold-based \emph{cardinality minimization problem} (CMP), a non-convex, non-smooth optimization problem.
	\item \emph{Solution Approaches.} We develop a continuous approximation of the threshold-based CMP formulation of SCED using the principles of difference-of-convex optimization (DCO). We express the differential thresholds of individual emergency operating zones as a difference of two piecewise convex functions. Such an approximation allows for the use of a difference-of-convex algorithm (DCA) to solve the program that exhibits desirable convergence and performance guarantees. Our approximation in conjunction with our DCO formulation for the SCED problem resolves the difficulty typically present in threshold-based CMP settings. We also develop a mixed-integer programming (MIP) formulation of the CMP that is suitable for use with off-the-shelf MIP solvers.
	\item \emph{Numerical experiments.} The computational experiments are the first of their kind that demonstrates the advantages of using the CMP formulation of SCED. The experiments reveal that the CMP-based model reduces the overall operating cost and avoids prolonged periods of high electricty prices during contingency events.
\end{enumerate}

The CMP captures the presence of emergency rating by employing a discrete indicator function called the $\ell_0$-function. Widely used in machine learning applications, 
$\ell_0$-function counts the number of nonzero-valued components, and performs variable selection to reconstruct an intrinsic sparse representation of the model. In practice, problems involving the $\ell_0$-function have been solved using reformulation and approximation techniques to avoid computational intractability coming from the discreteness and utilize existing optimization tools. The alternative formulations include MIP \cite{Bertsimas2016,Gmez2021}, complementarity constraints \cite{Burdakov2016,Feng2018,Xie2020}, and penalization methods. The complementarity approaches introduce auxiliary variables to formulate binary states of the variables (having zero and nonzero values) as orthogonality constraints. The penalization method introduces continuous $\ell_0$ surrogates to set the values of insignificant variables to zero through penalization. Existing surrogates include the convex $\ell_1$-norm \cite{Tibshirani1996} and nonconvex penalty functions \cite{fan2001variable,zhang2010nearly}. 

In this paper, we propose an approximate formulation that addresses the discrete property of CMP by introducing a continuous difference-of-convex surrogate function, then solve the problem using the DCA. The DCO approaches offer several advantages. The well-developed convex analysis supports the DCO and the associated algorithms are guaranteed to identify stationary points for nonconvex optimization problems. It has been shown that DCA produces a decreasing sequence of iterates that converge to a critical point where zero belongs to the subdifferential of the objective function, and to a directional stationary solution for specialized problems \cite{Pang2017}. The subproblems of DCA have been solved by utilizing efficient computational tools of convex programming. These reasons have kindled an interest in applying DCO for power systems operations problems. For instance, the optimal power flow (OPF) problem can be recast with constraints that are difference-of-convex quadratic functions. DCA was used in \cite{shi2017global, merkli2017fast} to solve the OPF problem for mesh networks and in \cite{Wei2017optimal} for radial networks. These works have demonstrated that the computational proficiency and scalability of DCO are comparable to alternative interior-point methods.

SCED is an important component used for contingency analysis. Contingency analysis ensures that the economic dispatch solution simultaneously meets a set of contingency scenarios corresponding to system component failures. Typically, operators consider sets of contingencies that include scenarios with at most one component failure (known as $N-1$ contingency analysis). A feasible solution to SCED must, upon a component failure, be able to redistribute the power flows across the system without overloading the remaining components. While the SCED model presented in this paper can be used for contingency analysis, we do not undertake such an endeavor. Instead, we focus on applying the model for corrective rescheduling following a contingency event.

The remainder of the paper is organized as follows. In \S\ref{sect:problemForm} we present the details of our CMP formulation of SCED. We first present the formulation in a single-period setting and later address the multiperiod setting through a rolling-horizon implementation. In \S\ref{sect:scedDCA} we develop a difference-of-convex approximation of the SCED formulation and present a solution method. We also present a MIP reformulation of SCED in this section. Finally, in \S\ref{sect:numericalExperiment}, we illustrate the performance of the proposed models and solution methods through numerical experiments. We present our conclusions and future research directions in \S\ref{sect:conclusions}.



\section{Problem Formulation} \label{sect:problemForm}
In this section we present a formulation of the economic dispatch problem that captures the operational as well as system reliability requirements. We will first present a single-period formulation of the dispatch problem, and then extend it to a multi-period setting. 

\subsection{Single-period Formulation}
We consider a day-ahead or an hour-ahead operations problem of a power system denoted by $(\set{B}, \set{L})$, where $\set{B}$ and $\set{L}$ are the sets of buses and lines, respectively. The goal of the operations problem is to minimize the total cost of operations. This problem is often stated with the following nominal (base-case) objective function:
\begin{align} \label{eq:operatingCostED}
    F_1(\x,\obs) := \sum_{g \in \set{G}} c_g p_g + \sum_{g \in \set{R}} s_g (\obs_g - p_g)_+ + \sum_{d \in \set{D}} s_d (\obs_d - p_d)_+. 
\end{align}
Here, $\x$ is a consolidated decision vector that includes decisions corresponding to generation quantities $(p_g)_{g \in \set{G}}$, utilized renewable generation $(p_g)_{g \in \set{R}}$, satisfied demand $(p_d)_{d \in \set{D}}$, line power flows $(f_{ij})_{(i,j) \in \set{L}}$, and any additional variables necessary to represent the power flows. The total cost of operations in \eqref{eq:operatingCostED} includes three terms corresponding to the total generation cost, opportunity cost associated with renewable curtailment, and load shedding penalties, respectively. We consider a setting where curtailing renewable generation is undesirable, and hence, the unutilized renewable generation $(\obs_g - p_g)_+ = \max\{\obs_g - p_g, 0\}$ is penalized at a rate of $s_g$ (\$/MWh) for all $g \in \set{R}$. Similarly, unmet demand $(\obs_d - p_d)_+ = \max\{\obs_d - p_d\}$ is also penalized at a rate of $s_d$ (\$/MWh) for all $d \in \set{D}$. From an operational point of view, $s_g$ can be interpreted as the cost of lost opportunity for renewable generators and $s_d$ as the loss of load penalty. 

The operations problem is considered in light of several requirements. These requirements are modeled as constraints in an optimization problem. The first set of constraints includes the following.
\begin{subequations} \label{eq:operationalConstraints}
\begin{align}
    & \sum_{j:(j,i) \in \set{L}} f_{ji} - \sum_{j:(i,j) \in \set{L}} f_{ij} + \sum_{g \in \set{G}_i \cup \set{R}_i} p_g \notag \\ 
    & \hspace{3cm}- \sum_{i \in \set{D}_i} p_d = 0 \qquad \forall i \in \set{B} \label{eq:flowBalance}\\
    & p_g^{\min} \leq p_g \leq p_g^{max} \qquad g \in \set{G} \label{eq:genCapacity}\\
    & 0 \leq p_g \leq \obs_g \qquad g \in \set{R} \label{eq:renewCapacity}\\
    & 0 \leq p_d \leq \obs_d \qquad d \in \set{D} \label{eq:demCapacity}\\
    & (f_{ij})_{(i,j) \in \set{L}} \in \mathfrak{F}. \label{eq:flowSet}
\end{align}
\end{subequations}
The flow balance equation in \eqref{eq:flowBalance} ensures that the net injection at all buses $\set{B}$ in the power network is zero. The constraint \eqref{eq:genCapacity} ensures that the generation amounts of all the generators that are operational, indexed by the set $\set{G}$, are within their respective minimum and maximum capacities. The constraint \eqref{eq:renewCapacity} restricts the amount of generation from renewable resources (indexed by the set $\set{R}$) utilized to be less than the total available generation $\obs_g$. Similarly, the demand met at load location $d$ is bounded from above by the actual demand $\obs_d$ for all $d \in \set{D}$. This is captured by constraint \eqref{eq:demCapacity}.

The set $\mathfrak{F}$ captures the physical requirements of the network. These include the active and the reactive flows, and the active and reactive power capacities for each line $(i,j) \in \set{L}$ in the power network. These requirements also include the lower and upper limits on the voltage magnitude and voltage phase angle ($V_i\angle\theta_i$) at each bus $i \in \set{B}$ of the power network. For instance, when one uses the linear direct-current approximation of the power flow, the set $\mathfrak{F}$ is a polyhedron given by
\begin{align} \label{eq:dcPowerFlow}
    \mathfrak{F} = \left\{ 
	\begin{array}{c}
	((f_{ij})_{(i,j) \in \set{L}} \\ 
	(V_i, \theta_i)_{i \in \set{B}})
	\end{array}	        
    \left \vert
\renewcommand{\arraystretch}{1.2}
\begin{array}{l}
f_{ij} = \frac{V_iV_j}{X_{ij}} (\theta_i - \theta_j) \quad \forall (i,j) \in \set{L} \\
f_{ij}^{\min} \leq f_{ij} \leq f_{ij}^{\max} \quad \forall (i,j) \in \set{L} \\
\theta_i^{\min} \leq \theta_i \leq \theta_i^{\max} \quad \forall i \in \set{B}
\end{array} \right. \right\}.
\end{align}
Alternatively, one could employ the recently developed convex relaxations of the power flow such as the quadratic convex relaxation \cite{coffrin2015qc}, the second-order conic relaxation \cite{kocuk2016strong}, and the semidefinite programming relaxation \cite{jabr2006radial}. When these convex relaxations are employed, the set $\mathfrak{F}$ reduces to a convex compact set. The solution method presented in \S\ref{sect:solutionMethod} is designed for convex feasible sets, and therefore, it is impervious to the particular modeling approach employed for the power flows. 

\begin{figure}[!h]
\centering
\begin{tikzpicture}

	\draw[red,fill=red] (-5.5,-0.5) rectangle (5.5,0.5);
	\draw[orange,fill=orange] (-5,-0.5) rectangle (5,0.5);
	\draw[yellow!60,fill=yellow!60] (-4,-0.5) rectangle (4,0.5);
	\draw[black!30!green,fill=black!30!green] (-2,-0.5) rectangle (2,0.5);
	
	\node at (0,0) {Normal};
	\node at (-3,0) {LTE};
	\node at (3,0) {LTE};
	\node at (4.5,0) {STE};
	\node at (-4.5,0) {STE};

	\draw[latex'-latex', thick] (-6,-0.5) -- (6,-0.5); \node at (6.1,-0.5) {$f$};
	\node at (0,-0.75) {0};
	\node at (2,-0.75) {$\zeta^n$};
	\node at (-2,-0.75) {$-\zeta^n$};
	\node at (4,-0.75) {$\zeta^\ell$};
	\node at (-4,-0.75) {$-\zeta^\ell$};
	\node at (5,-0.75) {$\zeta^s$};
	\node at (-5,-0.75) {$-\zeta^s$};

\end{tikzpicture}
\caption{Operating zones based on thermal ratings of lines} \label{fig:lineOperatingZones}
\end{figure}
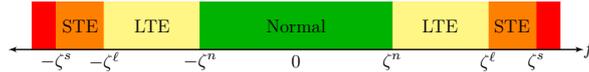

The power flow on transmission lines are additionally limited by the thermal ratings that are determined by the ISO/RTO's reliability standards. Under normal conditions, the system operates such that the transmission line and the corresponding equipment loading do not exceed a normal thermal rating. However, in the event of a contingency, the ISO operating procedures allow the use of less restrictive ratings for brief periods of time. These ratings are represented as operating zones marked by increasing levels of threshold. For instance, the California ISO imposes a $24$ hour (normal), $4$ hour (STE), and $15$ minute (LTE) ratings \cite{caiso2020establishing}. Similar operating practices are in place at other ISOs, albeit, the exact duration approved for operating in a zone and the threshold levels may differ based on rating methodologies. Along these practices, we adopt a similar three-tier operating zones for line $(i,j) \in \set{L}$ that are characterized by upper thresholds $\zeta_{ij}^n < \zeta_{ij}^\ell < \zeta_{ij}^s$.
\begin{itemize}
    \item \emph{Normal}: Flow is within the upper threshold of $\zeta^n$. Flow in this range, i.e., $|f_{ij}| \leq \zeta^n_{ij}$ is considered acceptable system performance. 
    \item \emph{Long-term emergency}: Flow is beyond $\zeta_{ij}^n$, but within the upper threshold of $\zeta_{ij}^\ell$, i.e., $\zeta_{ij}^n \leq |f_{ij}| \leq \zeta_{ij}^\ell$. Flow in this range for at most $4$ hours is acceptable.
    \item \emph{Short-term emergency}: Flow is beyond $\zeta_{ij}^\ell$, but within the threshold of $\zeta_{ij}^s$. This flow is captured by $\zeta_{ij}^\ell \leq |f_{ij}| \leq \zeta_{ij}^s$. Flow in this range for at most $15$ minutes is acceptable.
\end{itemize}
These operating zones are illustrated in Figure \ref{fig:lineOperatingZones}. 

Ordinarily, the SCED problem is formulated by restricting the flows to be within the normal operating zone. This is done by including constraints of the form: $|f_{ij}| \leq \zeta^n$ for all $(i,j) \in \set{L}$. Since the system can operate reliably even if the flow on transmission lines is outside the normal zone for short periods, accommodating the ability to operate outside the normal zone can help reduce the overall cost of operations. The perceived economic advantages are higher in power systems where a significant portion of the generation is from intermittent generators such as wind and solar. With large variability in the intermittent generation, the flows within the system exhibit large fluctuations, thereby increasing the probability of lines operating in emergency zones. A SCED model can accommodate the ability to operate outside the normal zone by simply relaxing the constraint $|f_{ij}| \leq \zeta^n$ and penalizing the amount of flow beyond the limit $\zeta^n$ for all $(i,j) \in \set{L}$. However, such an approach fails to account for the presence of different operating zones that dictate a power systems operator's response in case of a failure. Therefore, explicitly capturing the zone of operation of each line is desirable. Furthermore, since operating outside the normal zone increases the failure risk of a line, we desire to minimize the total number of lines operating in emergency zones. With this in mind, we present a SCED formulation which aims to minimize the number of lines operating in the emergency zones in addition to minimizing the operating cost.

In this regard, we consider that the flow beyond $\zeta_{ij}^s$ is unacceptable from a system reliability perspective. This restriction is imposed in our dispatch model as explicit constraints:
\begin{align} \label{eq:flowCapacity}
    |f_{ij}| \leq \zeta_{ij}^s \qquad \forall (i,j) \in \set{L}. 
\end{align}
To capture the number of lines operating in the emergency zones, we denote by $\set{E}^\ell$ and $\set{E}^s$ the sets of lines that are operating beyond the normal and LTE threshold values, respectively. These sets are defined as 
\begin{subequations}
\begin{align}
    & \set{E}^\ell := \{(i,j) \in \set{L}~|~|f_{ij}| \geq \zeta_{ij}^n\},~\text{and} \label{set:Eell} \\
    & \set{E}^s := \{(i,j) \in \set{L}~|~|f_{ij}| \geq \zeta_{ij}^\ell\}, \label{set:Es}
\end{align}
\end{subequations}
respectively. The additional system reliability requirement of minimizing the number of lines that do not operate in the normal zone is captured by the following function.
\begin{align} \label{eq:numLines}
    F_2(\x, \obs) = \gamma^\ell |\set{E}^\ell| + \gamma^s |\set{E}^s|.
\end{align}
Here, $\gamma^\ell, \gamma^s > 0$ are parameters that ensure that the STE zone is less desirable compared to LTE zone. Using the operational cost $F_1(\cdot)$ and system reliability objective $F_2(\cdot)$, defined in \eqref{eq:operatingCostED} and \eqref{eq:numLines}, respectively, the single period dispatch model can be stated as follows
\begin{align} \label{prob:scedCMP}
    \min~& F_1(\x,\obs) + F_2(\x, \obs) \\
    \text{subject to}~& \eqref{eq:operationalConstraints}, \eqref{eq:flowCapacity}. \notag
\end{align}
When flows are limited to operate within the normal threshold $\zeta^n$, the resulting SCED models are either convex or linear programs based on the approach adopted to model the power flows (i.e., the set $\set{F}$). However, in the presence of reliability objective $F_2(\cdot)$, the optimization program in \eqref{prob:scedCMP} is a CMP. The CMP is a non-smooth non-convex optimization problem, and therefore, directly solving \eqref{prob:scedCMP} is a computationally challenging undertaking. In order to tackle this difficulty, we present a computationally viable approximation of \eqref{prob:scedCMP} and a MIP reformulation in \S\ref{sect:scedDCA}.

\subsection{CMP-based Rolling-horizon SCED}\label{sect:rollingHorizon}
The system reliability requirements additionally mandate that an equipment return to normal operating zone within a fixed period of time. In order to incorporate these requirements, we extend the CMP-based SCED formulation to a multiperiod setting. 

\begin{figure*}[!t]
\centering
\begin{tikzpicture}[xscale=1.5,yscale=1.5]
	\node(xstart) at (-0.5,0){};
	\node(xend) at (2,0){};
	\node at (xend.east) {$|f_{ij}|$};
	\node(ystart) at (0,-0.5){};
	\node (yend) at (0,1.5) {};
	\node at (yend.north) {\footnotesize $\|\max\{|f_{ij}| - \zeta_{ij}^n,0\} \|_0$};
	
	\node [inner sep = 0](zero) at (0,0) {} (xstart) edge[->] (xend);
	\node [inner sep=0](zero) at (0,0) {} (ystart) edge[->] (yend);
	
	\draw (0.96,1) circle (0.04 cm); 
	\node [fill, circle, inner sep = 1.5pt](zero) at (0,0) {} ;
	\node [fill, circle, inner sep = 1.5pt](zero) at (1,0) {} ;
		
	\draw [line width=0.5mm] (1, 1) -- (1.8, 1) {};	
	\draw [line width=0.5mm] (0,0) -- (1,0) {};	
	
	\draw (1pt,1) -- (-1pt,1) node[anchor=west] {$1$};
	\draw (1,1pt) -- (1,-1pt) node[anchor=north] {\footnotesize $\zeta_{ij}^n$};
	
	\draw [decorate,decoration={brace,mirror,amplitude=5pt},xshift=0pt,yshift=0pt]
(0,-0.5) -- (0.95,-0.5) node [black,midway,xshift=0cm,yshift=-15pt] 
{\footnotesize Normal};
    \draw [decorate,decoration={brace,mirror,amplitude=5pt},xshift=0pt,yshift=0pt]
(1.05,-0.5) -- (2,-0.5) node [black,midway,xshift=0cm,yshift=-15pt] 
{\footnotesize LTE};
\end{tikzpicture}
    
\begin{tikzpicture}[xscale=1.5,yscale=1.5]
	\node(xstart) at (-0.5,0){};
	\node(xend) at (2,0){};
	\node at (xend.east) {$|f_{ij}|$};
	\node(ystart) at (0,-0.5){};
	\node (yend) at (0,1.5) {};
	\node at (yend.north) {\footnotesize $\phi_\varepsilon(f_{ij}; \zeta_{ij}^n)$};
	
	\node [inner sep = 0](zero) at (0,0) {} (xstart) edge[->] (xend);
	\node [inner sep=0](zero) at (0,0) {} (ystart) edge[->] (yend);
		
	\draw [line width=0.5mm] (0,0) -- (0.8,0) {};	
	\draw [line width=0.5mm] (0.8,0) -- (1.1,1) {};	
	\draw [line width=0.5mm] (1.1, 1) -- (1.8, 1) {};
	
	\draw (1pt,1) -- (-1pt,1) node[anchor=west] {$1$};
	\draw (0.8,1pt) -- (0.8 ,-1pt) node[anchor=north] {};
	\node at (0.75, -0.25) {\footnotesize $\zeta_{ij}^n$};
	\draw (1.1,0.5pt) -- (1.1 ,-0.5pt) node[anchor=north] {};
	\node at (1.2,-0.25) {\footnotesize $\zeta_{ij}^n + \varepsilon$};

    \node at (1.2, -0.9) {{\color{white} }};
\end{tikzpicture}

\begin{tikzpicture}[xscale=1.5,yscale=1.5]
	\node(xstart) at (-0.5,0){};
	\node(xend) at (2,0){};
	\node at (xend.east) {$|f_{ij}|$};
	\node(ystart) at (0,-0.5){};
	\node (yend) at (0,1.5) {};
	\node at (yend.north) {\footnotesize Components of $\phi_\varepsilon(f_{ij}; \zeta_{ij}^n)$};
	\node (g) at (1.1,1.17) {$g$};
	\node (h) at (1.4,1.2) {$h$};
	
	\node [inner sep = 0](zero) at (0,0) {} (xstart) edge[->] (xend);
	\node [inner sep=0](zero) at (0,0) {} (ystart) edge[->] (yend);
		
	\draw [line width=0.5mm] (0,0) -- (0.8,0) {};	
	\draw [line width=0.5mm] (0.8,0) -- (1.1,1) {};	
	\draw [dashed, line width=0.5mm] (0,0) -- (1.1,0) {};
	\draw [dashed, line width=0.5mm] (1.1,0) -- (1.4,1) {};
	
	\draw (1pt,1) -- (-1pt,1) node[anchor=west] {$1$};
	\draw (0.8,1pt) -- (0.8 ,-1pt) node[anchor=north] {};
	\node at (0.75, -0.25) {\footnotesize $\zeta_{ij}^n$};
	\draw (1.1,0.5pt) -- (1.1 ,-0.5pt) node[anchor=north] {};
	\node at (1.2,-0.25) {\footnotesize $\zeta_{ij}^n + \varepsilon$};
	
    \node at (1.2, -0.9) {{\color{white} }};
\end{tikzpicture}
\caption{Illustration of difference-of-convex approximation} \label{fig:dca}
\end{figure*}
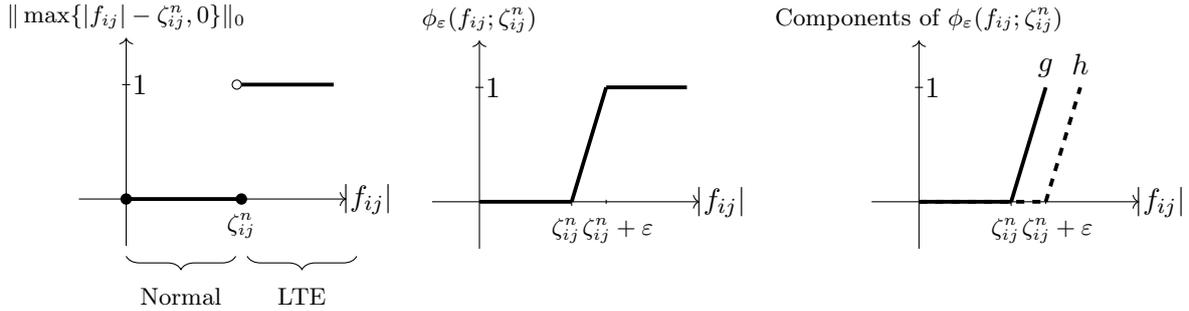
Denote by $\set{T} := \{1,\ldots,T\}$ the set of decision epochs in the multiperiod horizon. A time index $t$ will appear in the subscript for the applicable parameters and variables defined previously. Let $T^\ell$ and $T^s$ denote the acceptable number of time periods that a line can operate in LTE and STE zones, respectively. At any decision epoch $t \in \set{T}$ we denote the state of the system by a vector $s_t$ that includes the following components: (i) the generation level for conventional generators, $(p_{gt-1})_{g \in \set{G}}$, (ii) the flow on each line $(f_{ijt-1})_{(i,j)\in\set{L}}$, and (iii) the number of epochs since entering an emergency operating zone $z = \ell$ or $s$, denoted as $(\tau^z_{ijt-1})_{(i,j)\in\set{L}}$. A line operating in normal zone in time period $t$ will have $\tau_{ijt}$ set to zero.

Across multiple time periods, a SCED model must capture the generator ramp rate restrictions, by including constraints of the form:
\begin{align} \label{eq:ramping}
    \Delta_g^{\min} + p_{gt-1} \leq p_{gt} \leq \Delta_g^{\max} + p_{gt-1} \quad \forall g \in \set{G}, t \in \set{T},
\end{align}
where $\Delta^{\min}_g/\Delta^{\max}_g$ are minimum/maximum ramp limits. The system reliability requirements are captured by the following constraint for all $t \in \set{T}$:
\begin{align} \label{eq:maxTime}
    |f_{ijt}| \leq \zeta^n \mathbf{1}_{(\tau^\ell_{ijt-1} = T^\ell)} + \zeta^\ell \mathbf{1}_{(\tau^s_{ijt-1} = T^s)} \quad \forall (i,j) \in \set{L},
\end{align}
where $\mathbf{1}_{(\cdot)}$ is the indicator function. The above constraint enforces that the flow on line $(i,j)$ is within the LTE upper threshold, i.e., $|f_{ijt}| \leq \zeta^\ell$ if the time since entering the STE zone is equal to the acceptable amount $T^s$. Similarly, the flow is forced to return to normal zone once the acceptable amount of time for operating in LTE ($T^\ell$) is reached. 

Let $[t] \subset \set{T}$ denote the subset of decision epochs starting at time period $t$ and ending at time period $t+T^\prime$. That is, $[t] := t,t+1,\cdots,t+T^\prime$. For a given state $s_t$ and realization $\xi_t$, the multiperiod CMP-based SCED problem for time period $t$ is stated as
\begin{align} \label{prob:CMPmultiperiod}
    h_t(s_t, \obs_t) = \min~& \sum_{t^\prime \in [t]} \big(F_{1,{t^\prime}}(\x,\obs_{t^\prime}) + F_{2,{t^\prime}}(\x, \obs_{t^\prime}) \big),\\
    \text{subject to}~& \eqref{eq:operationalConstraints}, \eqref{eq:flowCapacity}, \eqref{eq:ramping}, \eqref{eq:maxTime} \quad \forall t^\prime \in [t]. \notag
\end{align}
For a given $s_t$, the right-hand side quantity in \eqref{eq:maxTime} can be computed easily for $t^\prime = t$ and the constraints appear as simple variable bounds.

In the rolling-horizon setting, an instance of the model in \eqref{prob:CMPmultiperiod} is solved for every time period $t$. While the optimal solution for the first time period ($t^\prime = t$) is implemented, the solutions for the remaining time periods $t^\prime \in [t]\setminus\{t\}$ are advisory in nature. These advisory decisions are overwritten by optimal solutions of instances solved in later time periods.

The optimal solution for $t^\prime = t$ is also used to update the state vector. In particular, the third component of the state vector is updated as follows:
\begin{align} \label{eq:updateStateVector}
    \tau_{ijt+1}^\ell =  \left\{
        \begin{array}{ll}
        \tau_{ijt}^\ell & \text{if } |f_{ijt}^\star| \leq \zeta_{ij}^n,\\
        \tau^\ell_{ijt}+1 & \text{if } |f_{ijt}^\star| > \zeta_{ij}^n
    \end{array} \right.
\end{align}
for all $(i,j) \in \set{L}$. The component $\tau_{ijt}^s$ is updated in a similar manner. A model instances is then setup using the updated state vector $s_{t+1}$ and solved for time period $(t+1)$. The procedure is continued until the end of the horizon. 

Finally, state variable usage also allows us to include the ability to monitor line cooling requirements. We define line cooling period, denoted by $T^c$, as the minimum number of time periods that the line is required to operate in normal zone after it is operated  in one of the emergency operating zone for the designated maximum amount of time. In other words, a line that operates in LTE zone for $T^\ell$  time periods or STE zone for $T^s$ time periods, must operate in normal zone for $T^c$ time periods. After $T^c$ time periods, we reset the state variables $\tau_{ijt+1}^\ell = \tau_{ijt+1}^s = 0$, for a line $(i,j) \in \set{L}$ that required cooling.

\section{Difference-of-Convex Approximation} \label{sect:scedDCA}
The CMP-based SCED models in \eqref{prob:scedCMP} and \eqref{prob:CMPmultiperiod} are non-smooth and non-convex optimization problems. The sets $\set{E}^\ell$ and $\set{E}^s$ in \eqref{prob:scedCMP} consist of all lines with flow beyond the desired capacity that the cardinality of the sets $|\set{E}^\ell|$ and $|\set{E}^s|$ represent the number of lines operating in the emergency zones. In this section, we present an exact formulation of such cardinalities, 
and introduce a computationally viable approximation of the quantities for the CMP formulation of SCED. 

We illustrate the principal technique on a line $(i,j)$ in the set $\set{E}^\ell$. For line $(i,j)$, it is not difficult to see that the line is operating in the LTE zone if $|f_{ij}| - \zeta_{ij}^n > 0$, or equivalently, if $\max \{ |f_{ij}| - \zeta_{ij}^n, 0 \} > 0$. For such a scenario, we can express the emergency operation of the line exactly by $\| \max \{ |f_{ij}| - \zeta_{ij}^n, 0 \} \|_0 = 1$, where the $\ell_0$-function, denoted by $\| \cdot \|_0$, is defined as $\| t \|_0 = 1$ if $t \neq 0$ and $\| t \|_0 = 0$ if $t=0$. The right-side of the equation becomes 0 if the line is operating under the normal zone. Hence the formulation indicates whether or not a flow exceeds the normal threshold $\zeta^n$; we refer the reader to the leftmost pane in Figure~\ref{fig:dca} for an illustration. A similar technique is applied to other lines in the set $\set{E}^\ell$ and the lines in the set $\set{E}^s$. The cardinality of the sets containing lines operating in the emergency zones can then be expressed as
\begin{subequations}  \label{def:exact_card}
\begin{align}
    |\set{E}^\ell| = \sum_{(i,j) \in \set{L}} \|\max\{|f_{ij}| - \zeta_{ij}^n,0\} \|_0\\
    |\set{E}^s| = \sum_{(i,j) \in \set{L}} \|\max\{|f_{ij}| - \zeta_{ij}^\ell,0\} \|_0.
\end{align}
\end{subequations}
We note that the notation $\| \cdot \|_0$ in \eqref{def:exact_card} is referred to as the $\ell_0$ norm for the special case of vector input. 
A common practice of solving the special case is replacing the function by continuous surrogates such as the $\ell_1$ norm \cite{Tibshirani1996} and nonconvex penalty functions \cite{fan2001variable, zhang2010nearly} that are sum of univariate symmetric folded concave functions.

As depicted in Figure~\ref{fig:dca}, the exact formulation of the cardinalities requires employing a discontinuous function, making any optimization problem involving such expressions a discrete problem. Extending the mentioned reformulation methods, we propose to approximate each summands in \eqref{def:exact_card} by a piecewise linear function to remove the discontinuity of the original formulation by connecting the two disjoint pieces of the function; the central pane of Figure~\ref{fig:dca} illustrates the continuous piecewise linear approximation for a single line. An advantage of using a surrogate, as opposed to directly solving a discrete optimization problem, is computational efficiency. 

Let us denote the approximation function for the summands in \eqref{def:exact_card} by $\phi_\varepsilon(\cdot; \cdot)$. The approximation function is defined for a variable (flow) and a parameter (threshold). For a given line $(i,j) \in \set{E}^\ell$ and the threshold $\zeta_{ij}^n$, we formally introduce the function: 
\begin{align} \label{def:phi}
    \phi_\varepsilon(f_{ij};\zeta_{ij}^n) 
    &= \max \left\{ \displaystyle{\frac{1}{\varepsilon}} ( |f_{ij}| - \zeta_{ij}^n ), 0 \right\} \notag \\ 
    & \hspace*{1cm } - \max \left\{ \displaystyle{\frac{1}{\varepsilon}} ( |f_{ij}| - \zeta_{ij}^n) - 1, 0 \right\} \\
    &=
    \left\{ 
    \begin{tabular}{ll}
        $0$ & if $|f_{ij}| \leq \zeta_{ij}^n$\\[0.5pc] \nonumber
        $\displaystyle{\frac{1}{\varepsilon}} ( f_{ij} - \zeta_{ij}^n )$ & if $\zeta_{ij}^n < |f_{ij}| \leq \zeta_{ij}^n + \varepsilon$\\[0.5pc] \nonumber
        $1$ & if $\zeta_{ij}^n + \varepsilon < |f_{ij}|$. \nonumber
    \end{tabular}
    \right.
\end{align}
The positive scalar $\varepsilon$ in the above definition is an approximation parameter that can be pre-selected or tuned in practice. The function is defined by three pieces -- the constant value of $0$ indicates that the flow is with the threshold, a value of $1$ indicates that the flow clearly exceeds the threshold, and a value between $0$ and $1$ indicates that the line just entered the emergency zone. Table \ref{tab:exactApproxForms} lists the exact and the approximate functions for all operating zones.
\begin{table}[!ht]
    \centering \renewcommand{\arraystretch}{1.5}
    \resizebox{0.49\textwidth}{!}{%
    \begin{tabular}{ccc} \hline \hline %
         Zone  & Exact formulation & Approximation \\ \hline \hline
         Normal & $\| \max \{ |f| - \zeta^n, 0 \} \|_0 = 0$ & $\phi_\varepsilon(f; \zeta^n) = 0$ \\[0.2pc] \hline
         \multirow{2}{*}{LTE} & $\| \max \{ |f| - \zeta^n, 0 \} \|_0 = 1$ & $0 < \phi_\varepsilon(f; \zeta^n) \leq 1$ \\
          & $\| \max \{ |f| - \zeta^\ell, 0 \} \|_0 = 0$ & $\phi_\varepsilon(f; \zeta^\ell) = 0$ \\[0.2pc] \hline
         \multirow{2}{*}{STE} & $\| \max \{ |f| - \zeta^n, 0 \} \|_0 = 1$ & $\phi_\varepsilon(f; \zeta^n) = 1$ \\
         & $\| \max \{ |f| - \zeta^\ell, 0 \} \|_0 = 1$ & $0 < \phi_\varepsilon(f; \zeta^\ell) \leq 1$
         \\[0.2pc] \hline \hline
    \end{tabular}}
    \caption{Exact and approximate formulations for different operating zones}
    \label{tab:exactApproxForms}
\end{table}
We note that each of the above approximation is a difference-of-convex function; a function $f(x)$ is a difference-of-convex function if there exist two convex functions $g(x)$ and $h(x)$ such that $f(x) = g(x) - h(x)$. To see this, consider the definition of $\phi_\varepsilon(\cdot;\cdot)$ given in \eqref{def:phi}. Each term in the right-hand side is a convex function since the max-operator preserves convexity given by the absolute value function. 

Applying the approximate function to all lines, the surrogates for the set  cardinalities in \eqref{def:exact_card} can be written as
\begin{align*} 
    |\set{E}^\ell| \approx \sum_{(i,j) \in \set{L}} \phi_\varepsilon(f_{ij}; \zeta_{ij}^n) \quad \text{and} \quad 
    |\set{E}^s| \approx \sum_{(i,j) \in \set{L}} \phi_\varepsilon(f_{ij}; \zeta_{ij}^\ell).
\end{align*}
Using the above construction, we introduce the approximation that reformulates \eqref{prob:scedCMP} by applying the surrogates of the cardinalities as:
\begin{align} \label{prob:aCMP}
    \min~& F_1(\x,\obs) \notag \\ & \hspace*{0.1cm}+ \gamma^\ell \sum_{(i,j) \in \set{L}} \phi_\varepsilon(f_{ij}; \zeta_{ij}^n) + \gamma^s \sum_{(i,j) \in \set{L}} \phi_\varepsilon(f_{ij}; \zeta_{ij}^\ell) \\
    &\text{subject to}~ \eqref{eq:operationalConstraints}, \eqref{eq:flowCapacity}. \notag
\end{align} 
We recall that $\gamma^\ell >0$ and $\gamma^s >0$ are the weighting parameters that control the number of lines operating in the emergency zones. We will refer to the SCED formulation in \eqref{prob:aCMP} as CMP with difference-of-convex approximation (CMP-DC). 

\subsection{Solution Method} \label{sect:solutionMethod}

\begin{algorithm}[!t]
    \caption{Difference-of-convex algorithm for \eqref{prob:aCMP}}
	\label{alg:DCA}
    \begin{algorithmic}[1]
	\State \textbf{Input:} Reliability parameters $\{\zeta_{ij}^n\}$ and $\{ \zeta_{ij}^\ell \}$ for all $(i,j) \in \set{L}$, and hyperparameters $\varepsilon, c, \sigma, \gamma^\ell,  \gamma^s > 0$.
	\State \textbf{Initialization:} Set $k=0$. $f^k = (f_{ij}^k)_{(i,j) \in \mathcal{L}}$ \;
	\While {$|F(\x^{k-1},\xi^{k-1}) - F(\x^k, \xi^k)| / |F(\x^k,\xi^k)|   > \sigma$}
	\State Compute the subgradient:
	    \begin{align}
	    v_{ij}^{k,n} = \left\{ 
	    \begin{array}{ll}
	        -\frac{1}{\varepsilon} & \text{if } f_{ij}^k < \zeta_{ij}^n - \varepsilon \\
	        \left[ -\frac{1}{\varepsilon},0 \right] & \text{if } f_{ij}^k = \zeta_{ij}^n - \varepsilon \\
	        0 & \text{if }\zeta_{ij}^n - \varepsilon < f_{ij}^k < \zeta_{ij}^n + \varepsilon \\
	        \left[ 0, \frac{1}{\varepsilon} \right] & \text{if } f_{ij}^k = \zeta_{ij}^n + \varepsilon \\
	        \frac{1}{\varepsilon} & \text{if } \zeta_{ij}^n + \varepsilon < f_{ij}^k. \\
	    \end{array}
 	    \right.
 	    \end{align}
	    \State Solve the subproblem
	    \begin{align} \label{eq:dcaSubproblem}
	    \min~& F(\x, \xi) \triangleq F_1(\x,\obs) + \frac{c}{2} \| \f - \f^k \|_2^2 \notag \\ 
		& \hspace*{1cm} 
		+ \gamma^\ell \sum_{(i,j) \in \set{L}} \left\{ g(f_{ij}; \zeta_{ij}^n) - v_{ij}^{k,n} \, f_{ij} \right\} \notag \\ 
		&\hspace*{1cm} 
		+ \gamma^s \sum_{(i,j) \in \set{L}} \left\{ g(f_{ij}; \zeta_{ij}^s) - v_{ij}^{k,\ell} \, f_{ij} \right\}  \notag \\
		\text{subject to}~& \eqref{eq:operationalConstraints}, \eqref{eq:flowCapacity};
        \end{align} 
	    \State Let $\x^k$ denote the optimal solution of \eqref{eq:dcaSubproblem} with vector $\f^k = (f_{ij}^k)_{(i,j) \in \set{L}}$ corresponding to the flow variables.
 	    \State $k = k + 1$ \;
    \EndWhile 
	\State \textbf{Output:} The optimal solution $\x^* = \x^k.$
\end{algorithmic}
\end{algorithm}
The problem \eqref{prob:aCMP} has a difference-of-convex objective function and convex constraints. A popular approach to solve a problem with such structure is to apply DCA. Introduced by Le Thi and Pham Dinh \cite{PhamHoiAn97}, the DCA iteratively solves a convex program which is given by a local approximation of the objective function. At each iteration, the algorithm linearizes concave components of the objective function using the current point, and solves the resulting convex problem producing decreasing sequence of iterates. 

Employing DCA to solve the problem \eqref{prob:aCMP}, we linearize each concave part of the objective given by  $\phi_\varepsilon(\cdot; \cdot)$. For ease of presentation of the algorithm, let us denote $g$ and $h$ for the two functions shown in the definition of $\phi(\cdot;\cdot)$ in \eqref{def:phi}, i.e., 
\begin{align} \label{def:g}
    &g(f_{ij};\zeta_{ij}^n) = \max \left\{ \displaystyle{\frac{1}{\varepsilon}} ( |f_{ij}| - \zeta_{ij}^n ), 0 \right\} \\
    &h(f_{ij};\zeta_{ij}^n) = \max \left\{ \displaystyle{\frac{1}{\varepsilon}} ( |f_{ij}| - \zeta_{ij}^n )-1, 0 \right\}. \nonumber 
\end{align}
Given a current point $f_{ij}^k$ at the $k$-th iteration, we approximate the latter function using its subgradient. 
\begin{align*}
    h(f_{ij};\zeta_{ij}^n) 
    \approx
    h(f_{ij}^k;\zeta_{ij}^n) 
    + v_{ij}^k (f_{ij} - f_{ij}^k)
\end{align*}
where $v_{ij}^{k,n} \in \partial h(f_{ij}^k,\zeta_{ij}^n)$. We note that the approximation shown on the right-hand side is a linear function with some constant terms. Since minimizing without constant terms does not affect the solution of the problem, we discard the constant terms in the algorithm. Incorporating the linear approximation, we present Algorithm~\ref{alg:DCA} to solve the problem \eqref{prob:aCMP}. The algorithm computes iterates with decreasing objective values. We stop the algorithm when the relative difference between two consecutive objective values is within a prescribed value as shown in \cite{Lipp2015} and \cite{JaraMoroni2017}.

\subsection{Mixed-Integer Programming Formulation} \label{sec:mip}
The optimization problem in \eqref{prob:scedCMP} also admits a MIP reformulation. To present this reformulation, we use binary decision variables to identify the lines in each of the emergency zones. Let $z_{ij}^\ell$ and $z_{ij}^s$ denote the binary variables that take a value of one if the line operates in the LTE and STE zones, respectively, and zero otherwise. \change{The single-period MIP reformulation is stated as}
\begin{subequations}\label{prob:mip}
\begin{align}
    \min~&F_1(\x,\obs) + \sum_{(i,j) \in \set{L}} \big(\gamma^\ell z_{ij}^\ell + (\gamma^\ell + \gamma^s) z_{ij}^s\big) \\
    \text{s.t.}~& \eqref{eq:operationalConstraints}, \eqref{eq:ramping}, \notag \\
    & |f_{ij}| \leq \zeta_{ij}^{\ell} z_{ij}^\ell + \zeta_{ij}^sz_{ij}^s, \qquad \forall (i,j) \in \set{L}, \label{eq:mip_lineLimit} \\
    & z_{ij}^\ell, z_{ij}^s \in \{0,1\} \quad (i,j) \in \set{L}. 
\end{align}
\end{subequations}
Notice that the above formulation includes two binary variable for every line. The constraint \eqref{eq:mip_lineLimit} enforces limits on capacity on power flows through line $(i,j) \in \set{L}$. For instance, when $z_{ij}^\ell = 1$, the line capacity is restricted to the LTE limit $\zeta_{ij}^\ell$. Similarly, $z_{ij}^s = 1$ relaxes the capacity to the STE limit $\zeta_{ij}^s$. We will refer to the formulation in \eqref{prob:mip} as CMP-MIP. Note that the MIP reformulation is a mixed-integer linear program when direct-current approximation is employed in the description of $\mathfrak{F}$ used in \eqref{eq:operationalConstraints}. When convex relaxations of optimal power flow are employed, the resulting optimization problem is a mixed-integer nonlinear program. \change{Further, the multiperiod mixed-integer extension of \eqref{prob:mip} includes the ramping constraints \eqref{eq:ramping} and the system reliability requirements \eqref{eq:maxTime}. The latter can be implemented efficiently similar to the minimum generator up-time and down-time constraints in a unit commitment problem (see \cite{wu2019security, khodaei2010transmission}, for example). The resulting model is a mixed-integer variant of \eqref{prob:CMPmultiperiod}.} 

\section{Numerical Experiments} \label{sect:numericalExperiment}
In this section we report the results from the numerical experiments with the CMP formulations of SCED. We use three test power systems available in the literature for our numerical experiments. The first test system is an updated version of the RTS-96 test system \cite{RTS-GMLC}. The test system comprises of $73$ buses, $108$ lines, and $158$ generators. The other two test systems are IEEE 118 and IEEE 300 from \cite{PSTCA}. The IEEE 118 test system has $118$ buses, $179$ lines, and $54$ generators. The corresponding numbers for IEEE 300 test system are $300$, $409$, and $57$, respectively. The experiments were conducted on a computer using a 2.7 GHz Intel Core i5 processor with 8GB of RAM, running macOS Sierra version 10.12.6. Gurobi Optimizer version 9.0.0 through CVX was used in MATLAB. 
 
We consider a time resolution of $15$ minutes that corresponds to the acceptable time limits for the LTE rating. Following this choice, the parameters  $T^\ell = 16$ and $T^s = 1$ (number of $15$-minute time intervals in $4$ hours). In our experiments, we consider a rolling-horizon length of $24$ hours resulting in a total of $T = 96$ decision epochs over the horizon. Individual optimization problems solved in our rolling-horizon setting use $T^\prime = 0$. This is consistent with the practice at ISOs where only here-and-now problems are solved for each $t \in \set{T}$ (see for e.g., CAISO operations manual \cite{CAISO2017b}).

The original load data that has a time resolution of $5$-minutes was transformed into $15$-minute intervals by averaging over corresponding three time intervals. Additionally, the load time series was multiplied by a constant factor to emulate a contingency scenario in the system. For our experiments, we consider scenarios where the system is stressed for a prolonged period of time (multiple hours). We assume that the set of generators is committed and no additional resources can be brought online in real time. This assumption is made to focus the experiments on illustrating the ability of the CMP to efficiently utilize the available line capacities. The generation cost data is included in the data set. Estimating the opportunity cost is challenging in general \cite{Wellenius2018challenges}. These costs range between $\$50 - \$500$/MWh (see \cite{Lori2014wind} for lost opportunity cost payments at Pennsylvania-New Jersey-Maryland (PJM) ISO). The value of lost-load varies between $\$0-\change{\$42000}$ based on the nature of load (residential v. industrial) and the system operator (see \cite{Frayer2013}). In our experiments we set $s_g = \$300$ and $s_d = \$1000$. These penalties are applied uniformly for all the generators and loads, respectively. 

In our experiments, we use two additional models to serve as benchmark to the CMP models. The first model is the ``strict'' model that does not use cardinality minimization and imposes all the lines to operate within the normal zone for every time period. This model is stated as
\begin{align} \label{eq:strictModel}
    \min~\{F_1(\x,\obs)~|~\eqref{eq:operationalConstraints}, \eqref{eq:ramping}, |f_{ij}| \leq \zeta_{ij}^n\}.
\end{align}
The second model is a ``relaxed'' model that disregards the thermal rating of the lines and imposes the lines to operate within their capacity limits for every time period. This model is stated as
\begin{align} \label{eq:relaxedModel}
    \min~\{F_1(\x,\obs)~|~\eqref{eq:operationalConstraints}, \eqref{eq:ramping}, f_{ij}^{\min} \leq f_{ij} \leq f_{ij}^{\max}\}.
\end{align}
Since the CMP models (CMP-DC in \eqref{prob:aCMP} and CMP-MIP in \eqref{prob:mip}), the strict model \eqref{eq:strictModel}, and the relaxed model \eqref{eq:relaxedModel} adhere to the physical requirements and limitations, viz., \eqref{eq:operationalConstraints} and \eqref{eq:ramping}, their total operating costs are comparable. In our experiments, we use the direct-current approximation of power flows in the description of $\mathfrak{F}$ in \eqref{eq:operationalConstraints}.

Before we present the results that compare the above models with the CMP formulation of SCED, we will present the steps undertaken to identify the hyperparameters used in the difference-of-convex approximation.

\subsection{Hyperparameter Tuning}
Recall that the difference-of-convex Algorithm \ref{alg:DCA} requires hyperparameters $\varepsilon$, $\gamma^\ell$, and $\gamma^s$ as input. A 3-dimensional grid search between $\varepsilon$, $\gamma^\ell$, and $\gamma^s$ was performed to choose hyperparameter values for the CMP-DC formulation. Five values of $\varepsilon$ were compared across $10$ different values for both $\gamma^\ell$ and $\gamma^s$. The $5$ tested values for $\varepsilon$ were powers of 10 evenly spaced from $10^{-4}$ to $10^{0}$. Tested values for $\gamma^\ell$, and $\gamma^s$ ranged from $0.1$ to $1.0$ in increments of $0.1$. For each combination of $\varepsilon$, $\gamma^\ell$, and $\gamma^s$, the total operating cost that includes the cost of generation plus the shedding costs, detailed in equation \eqref{eq:operatingCostED}, was recorded. The number of power lines within each operating zones at the end of each 15-minute interval were also recorded. 

Our analysis reveals a tradeoff between number of lines in the normal zone and total operating cost. When $\varepsilon=10^{-4}$, the total operating cost greatly increases, but there are more power lines operating withing the desirable thermal rating. We see the operating cost and subsequent number of lines operating in desirable thermal ratings both decrease as $\varepsilon$ increases. The harsher penalties associated with small $\varepsilon$ values leads to more demand shedding, as the penalty for each threshold violation begin to outweigh shedding penalties. Because the overall objective of the SCED problem is to minimize the total cost while still satisfying physical and thermal rating requirements, hyperparameter values associated with low costs are chosen for our experiments. The hyperparameter values for the CMP-DC formulation subsequently discussed uses $\varepsilon=10^{-1}$, $\gamma^\ell = 0.5$,  $\gamma^s = 0.5$, and $c = 0$.

\begin{table*}[!t]
    \centering \renewcommand{\arraystretch}{1.5}
    \resizebox{0.95\textwidth}{!}{
    \begin{tabular}{|l||cccc|cccc|cccc|} \hline 
         \multirow{2}{*}{Day} & \multicolumn{4}{c|}{Total Cost for the Day (in $10^8$ \$)} & \multicolumn{4}{c|}{Total Units of Demand Shed (in $10^5$ MW)} & \multicolumn{4}{c|}{Average \# Lines in Normal Zone} \\
        &CMP-DC & CMP-MIP & Strict & Relaxed & CMP-DC & MIP-MIP & Strict & Relaxed & CMP-DC & CMP-MIP & Strict & Relaxed  \\\hline
        Winter-1 & $1.613$ & $1.620$ & $1.712$ & $1.339$ & $1.337$ & $1.343$ & $1.430 $& $1.066 $ & $102.46$ & $103.05$ & $108$ & $95.36$ \\
        Winter-2 & $1.514$ & $1.523$ & $1.607$ & $1.251$ & $1.242$ & $1.247$ & $1.326 $ & $0.982$  & $102.635$ & $102.938$ & $108$  & $95.50$\\
        Spring-1 & $1.343$ & $1.348$ & $1.414$ & $1.105$  & $1.074$ & $1.078$ & $1.137 $ &$0.842$  & $103.15$ & $103.65$ & $108$ & $ 96.92$\\
        Spring-2 & $1.274$ & $1.278$ & $1.348$ & $1.042$ &  $1.006$ & $1.006$ & $1.071 $ & $0.779$ & $103.05$ & $103.82$ & $108$& $96.99$\\
        Summer-1 & $7.201$ & $7.201$ & $7.284$ & $7.055$& $6.933$ & $6.933$ & $7.015 $& $6.788 $ & $97.45$ & $102.86$ & $108$& $96.10$\\
        Summer-2 & $7.627$ & $7.628$ & $7.712$ & $7.478 $ & $7.360$ & $7.360$ & $7.443 $& $7.211$ & $97.14$ & $103.12$ & $108$& $97.01$\\
        Fall-1 & $4.302$ & $4.310$ & $4.404$& $4.147$ & $4.023$ & $4.029$ & $4.122 $& $3.871$ & $98.70$ & $103.91$ & $108$& $98.13$\\
        Fall-2 & $3.955$ & $3.961$ & $4.058$& $3.387$ & $3.677$ & $3.680$ & $3.776 $ &$3.510 $& $98.83$ & $103.46$ & $108$& $97.91$\\
        [.29pc]\hline
    \end{tabular}}
    \caption{Comparison across different models for various days on RTS-96 test system} \vspace{-0.5cm}
    \label{tab:DifferentDays}
\end{table*}

\subsection{Model Comparison} \label{subsec:modelComparison}

We applied Algorithm \ref{alg:DCA} to solve CMP-DC and used Gurobi's MIP solver to solve CMP-MIP. We compare the results obtained from the CMP models with the strict and relaxed models. For this comparison, we performed the experiment on eight different contingency scenarios, with two days corresponding to each of the four seasons in the RTS-96 test system. The results are shown in Table \ref{tab:DifferentDays}. The table shows the total operating cost, the units of demand shed, and the average number of lines operating within the normal zone for all the models. The total operating costs include the generation and shedding costs (if any).

The table shows that operating costs are the highest during summer contingency events. The high contingency summer days also result in the lowest number of lines operating in the normal zone for the CMP-DC ($97.45$ and $97.14$ out of $108$, respectively). When compared to the strict model, the CMP models that accommodate the flexibility of operating the lines outside the normal zone result in lower total operating costs. For instance, The operating costs resulting from CMP-DC formulation were lower by as much as $6.14\%$ (Winter-1) when compared to the strict model. The added flexibility also results in reducing the total demand shed (between $2-7\%$).  

The relaxed model, as expected, achieves the lowest total operating costs in all instances by overusing emergency thermal ratings. This is observed even on the winter and spring days that have lower demand. However, the lower operating costs come with an increased number of lines operating in the emergency zones (see the last column in Table \ref{tab:DifferentDays} that shows the average number of lines operating in the normal zone).
The results in Table \ref{tab:DifferentDays} also show that the CMP-DC model results in a lower total operating cost uniformly across all scenarios compared to the CMP-MIP model. The largest difference ($0.59\%$) between the two models was observed on day Winter-2. 

In order to study the scalability of the CMP models, we performed experiments with three test power systems with varying sizes. Table \ref{tab:SystemSizes} summarizes the results of this study. The table shows the total operating cost and computational time (per instance) for Winter-1 of RTS-96, and an arbitrarily chosen day of IEEE 118 and IEEE 300 systems. 
\begin{table}[!h] \renewcommand{\arraystretch}{1.3} \centering
    \begin{tabular}{|l|cc|cc|} \hline 
         \multirow{2}{*}{System} & \multicolumn{2}{c|}{Total Operating Cost} & \multicolumn{2}{c|}{Computation Time} \\         \cline{2-5}
        &CMP-DC & CMP-MIP & CMP-DC & CMP-MIP  \\\hline
        RTS-96 & $1.613\times 10^8$ & $1.620\times 10^8$ & $56.95 $ sec & $26.51$ sec  \\
        IEEE 118 & $7.880\times 10^6$ & $7.880\times 10^6$ & $93.34$ sec & $59.67$ sec\\
        IEEE 300 & $3.786\times 10^7$ & $3.795\times 10^7$ & $22.25$ min & $20.53$ min \\ \hline
    \end{tabular}
    \caption{Comparison between CMP-DC and CMP-MIP models for different test systems}  
    \label{tab:SystemSizes}
\end{table}
We see the CMP-DC model results in equal or lower costs when compared to the CMP-MIP model for all the systems tested. The relative behavior of strict and relaxed models on the larger test systems was similar to those reported in Table \ref{tab:DifferentDays}.

While the CMP-DC's computation time is larger than the CMP-MIP, we note that the computational performances become more comparable as the system size increases. The CMP-DC computation time is more than two folds higher for RTS-96, this reduces to $8.38\%$ for the larger IEEE 300 test system. The higher computational time for CMP-DC model is attributed to the time used to setup the problem on the CVX solver. On the other hand, the setup time on the commercial Gurobi solver was minimal.


\subsection{Cost Performance and Prices}

\begin{figure*}[!t]
    \centering
    \subfloat[Number of lines in each operating zone for CMP-DC]{\includegraphics[width=0.48\textwidth]{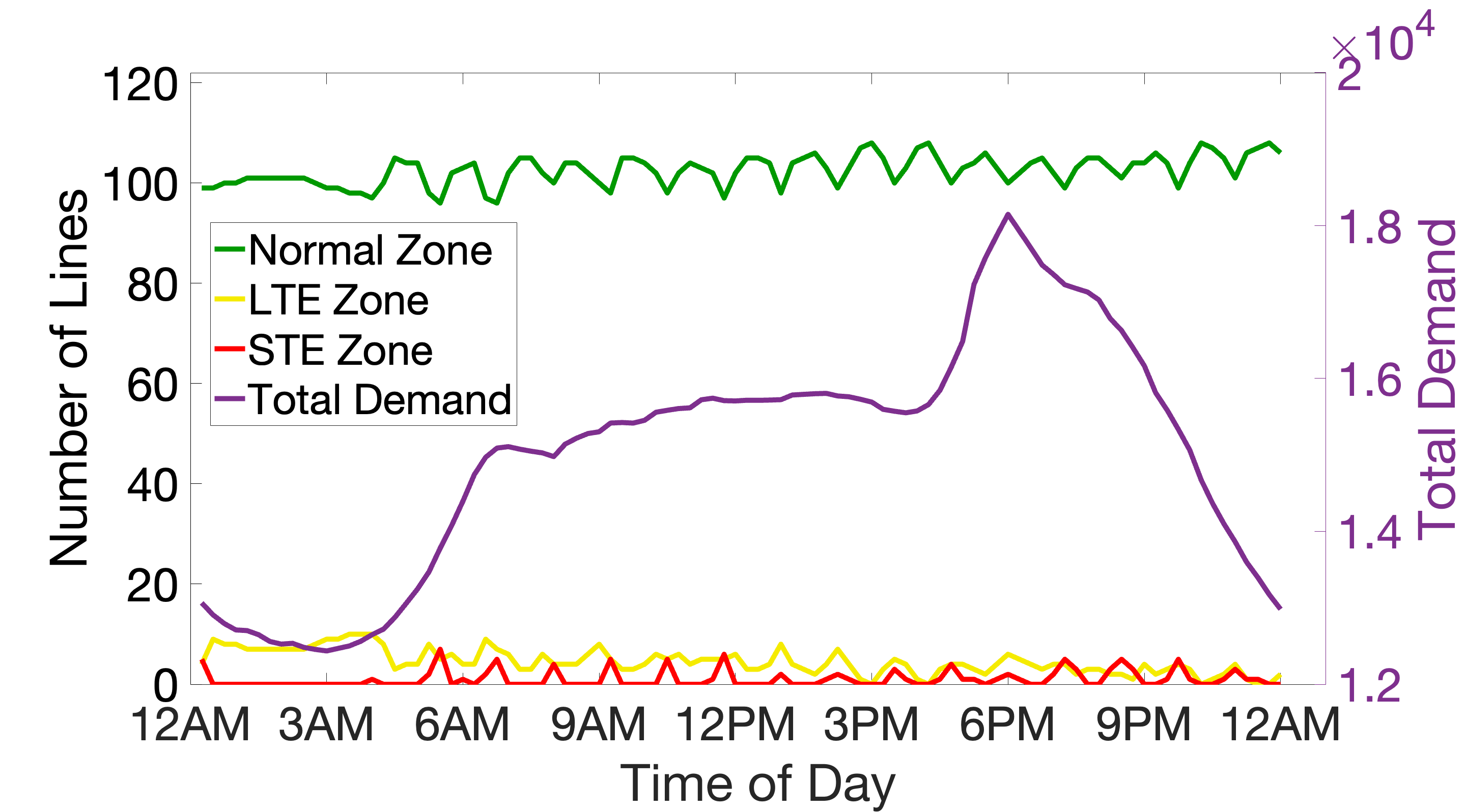}%
    \label{fig:LineDistribution}}
    \hfil
    \subfloat[Total operating cost comparison]{\includegraphics[width=0.48\textwidth]{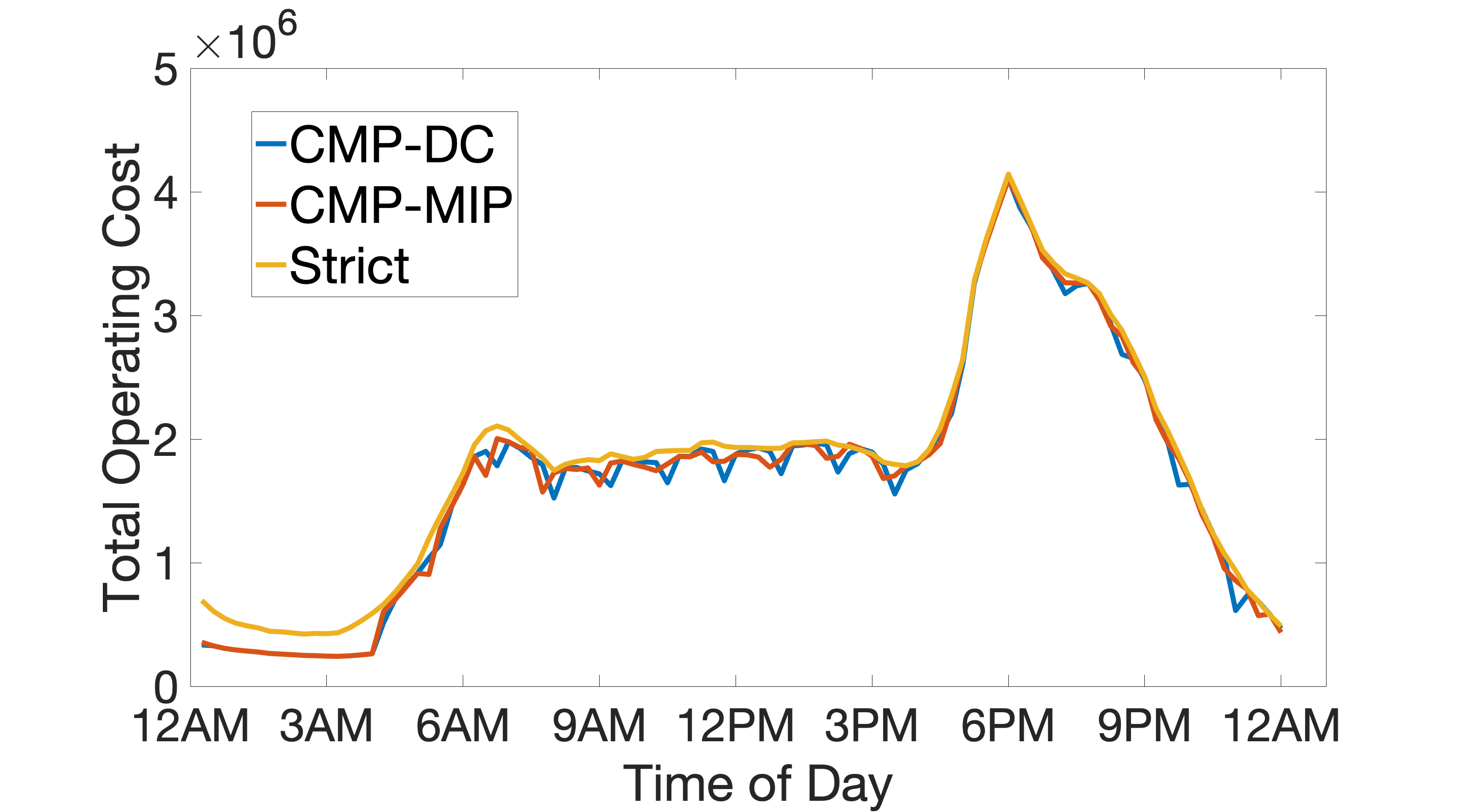}%
    \label{fig:OperatingCost}}
    \caption{Performance of the CMP models with cooling on RTS-96 system. } \label{fig:CMPperformace}
\end{figure*}

Figure \ref{fig:CMPperformace} shows more detailed performance of the CMP models for Winter-1 of RTS-96 test system in terms of the number of lines in each of the operating zones and the total cost of operations. In this study, the CMP models additionally requires each line that exits an emergency zone stay in the normal zone for the duration of line cooling period. The line cooling period was arbitrarily set to $60$ minutes (four time periods). 

Figure \ref{fig:LineDistribution} shows the number of lines operating in different zones when CMP-DC model is used for RTS-96 system. The behavior of CMP-MIP model was similar. We see that the model utilizes its flexibility as early as possible. The constraints \eqref{eq:maxTime} ensure that the time duration that an individual line operates outside the normal zone without cooling is within the acceptable reliability parameters. Therefore, the number of lines operating outside the normal zone decreases with time. However, once the lines have met the cooling period requirement, they once again operate in an emergency zone. To illustrate this behavior, note that there are $6$ lines in LTE zone at 11:45 A.M. that reduces to $3$ by 12:00 P.M. The number of lines in LTE again reaches $6$ at 12:45 P.M.  Therefore, we see an oscillating pattern in the number of lines where the period of oscillation aligns with the line cooling period $T^c$. The added flexibility of allowing the lines to operate outside the normal zone results in efficient utilization of resources to address contingency. This is reflected in the lower operating costs when compared to the strict model that lacks this flexibility as seen in Figure \ref{fig:OperatingCost}. The oscillating behavior is also evident in the total operating costs.

\begin{figure}[!h] \centering
    \includegraphics[width=0.43\textwidth]{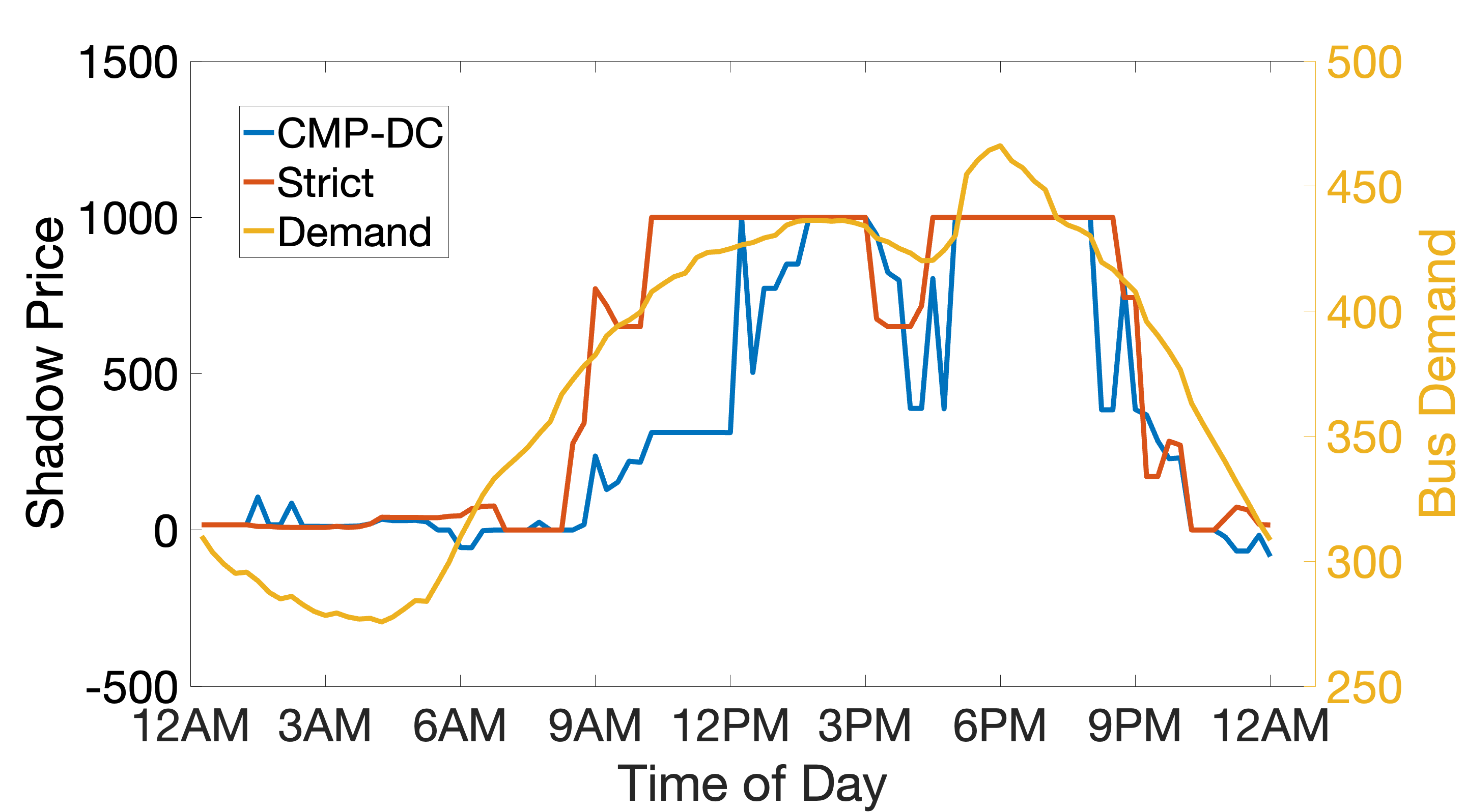}%
    \caption{Location marginal price at Bus-57}
    \label{fig:shadowPrice}  
\end{figure}

Figure \ref{fig:shadowPrice} shows the location marginal price obtained from the strict and CMP-DC models at Bus 57 which corresponds to the 75th percentile of demand in the network. The figure also shows the system demand over the $24$ hour horizon under the scenario corresponding to Winter-1. The location marginal prices are the optimal dual solution (shadow prices) associated with the flow balance equation \eqref{eq:flowBalance}. The strict model results in a scarcity price of $\$1000$ for prolonged periods of time, once between 9:30 A.M. -- 3:15 P.M. and again between 4:30 P.M. -- 9:00 P.M. On the other hand, the CMP-DC model reduces the intervals of scarcity prices during a contingency. Note the CMP-MIP model does not provide a direct means to compute the location marginal prices. This marks a critical distinction between CMP-DC and CMP-MIP models that is of significance from a systems operators' perspective.

\subsection{Effect of Line Cooling Period}
The line cooling period has a direct impact on the degree of flexibility offered by the CMP formulation of SCED. To examine this impact, we compare two settings that differ only in the line cooling period. In setting-I, the line cooling period is set to $60$ minutes. In setting-II, we impose that the lines can be in LTE/STE zones for their respective maximum periods only once in $\set{T}$. The operating cost in settings I and II are $1.613\times 10^8$ and $1.656\times 10^8$, respectively.

\begin{figure*}[!t]
    \centering
    \subfloat[Setting I: each line has cooling period of 60 minutes] {\includegraphics[width=0.45\textwidth]{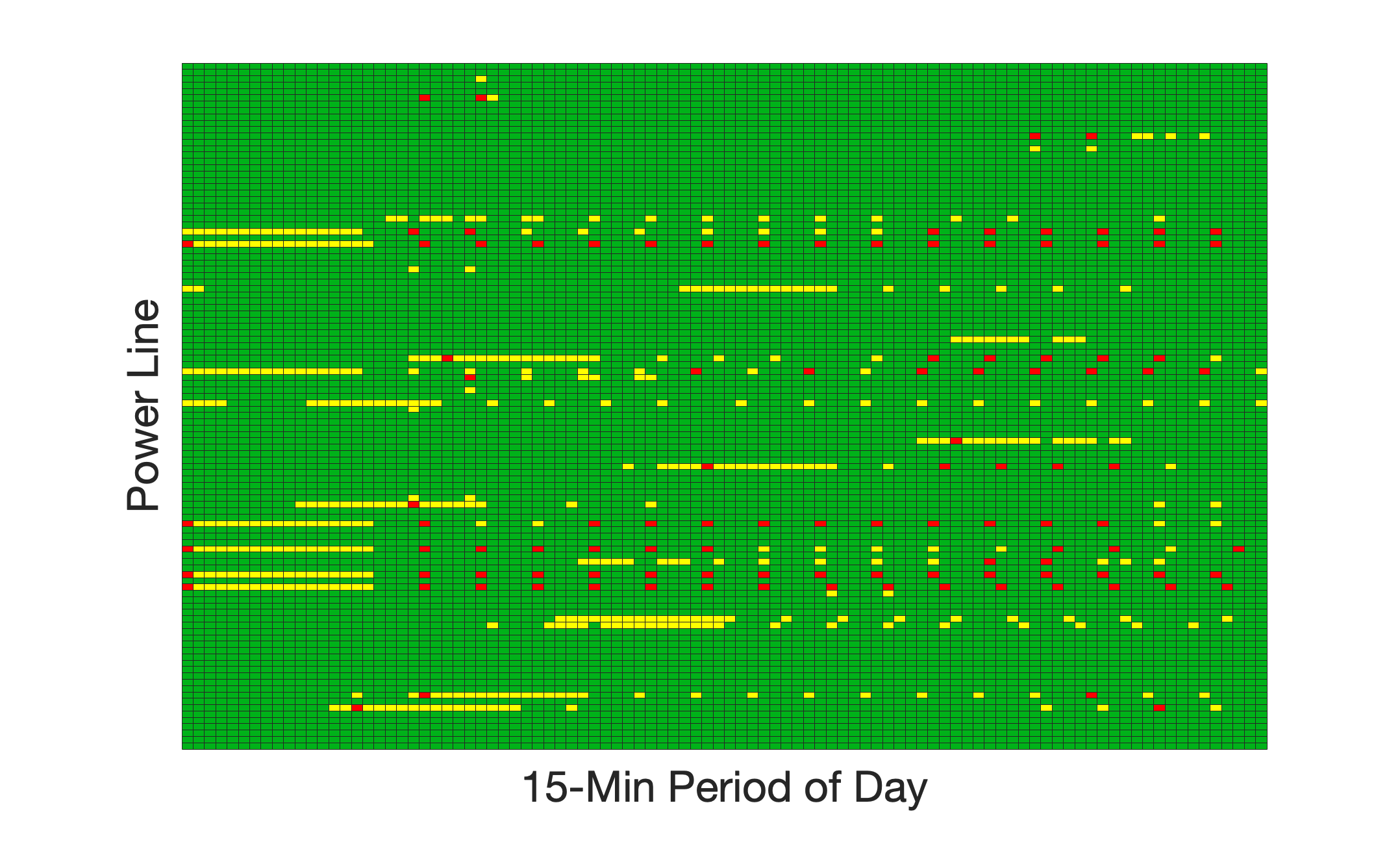}%
    \label{fig:WithCool}}
    \hfil
    \subfloat[Setting II: a line can stay in LTE(STE) zone for up to 16(1) time periods]{\includegraphics[width=0.45\textwidth]{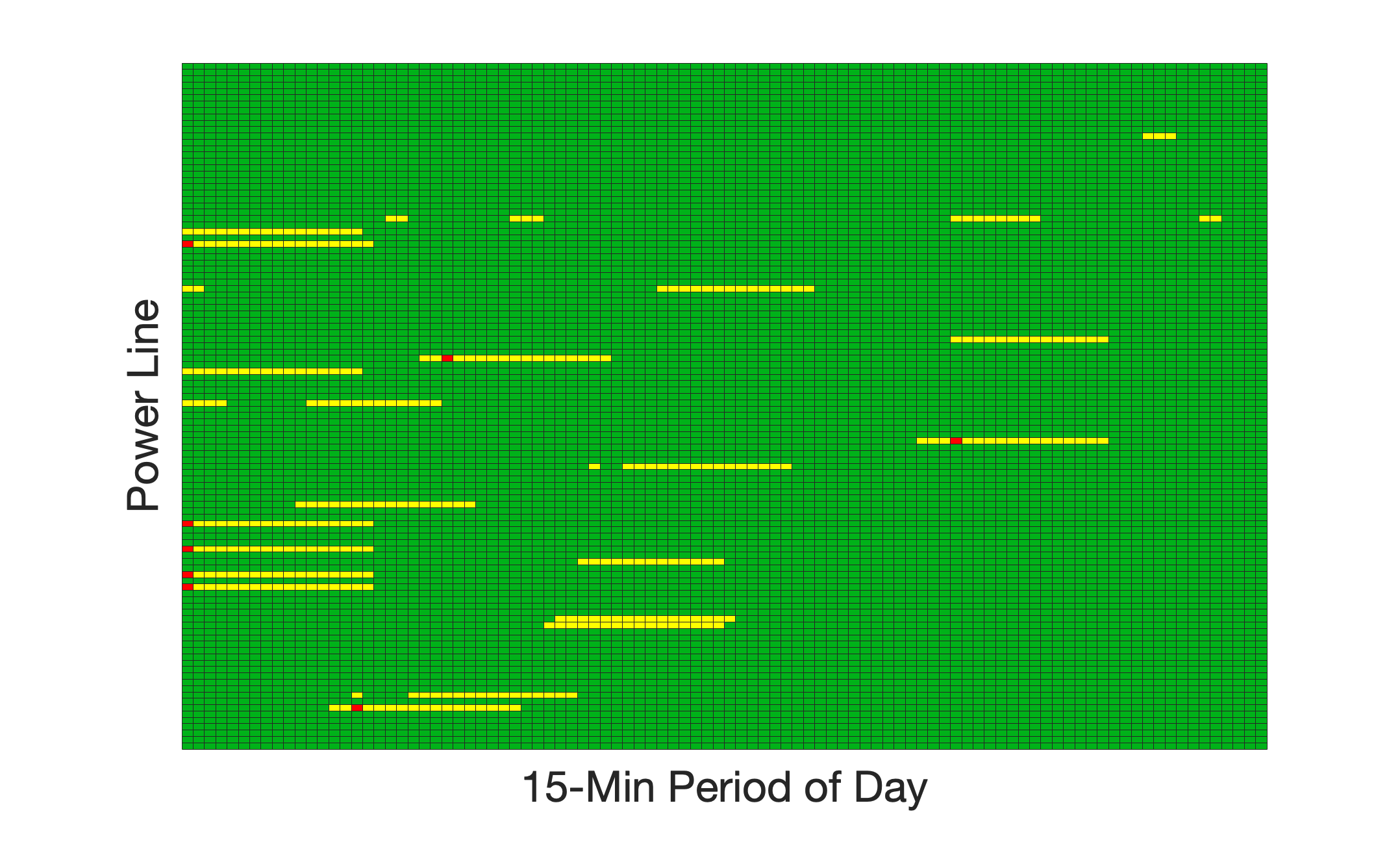}%
    \label{fig:WoutCool}}
    \caption{Heatmap of each power line for RTS96 System showing their operating zones (Winter-1). \change{The colors green, yellow, and red indicate that the line operates in normal, LTE, and STE zones respectively.}}
    \vspace{-0.5cm}
    \label{fig:Heatmaps}
\end{figure*}

We examine the activity of individual power lines through heatmaps presented in Figure \ref{fig:Heatmaps}. Each row of grid represents a power line, and each column represents a time period in $\set{T}$. A green square indicates that the line operates within the normal zone, i.e., $ |f_{ij}| \leq \zeta^n$, while yellow and red squares indicate the line operates in the LTE and STE zones, respectively. In setting II, we see that the model utilizes much of its flexibility for critical lines early in the day and later with other power lines as demand increases. On the other hand, in setting I, the model allows the same, critical lines to re-enter emergency zone - after the line cooling period requirement is met - as long as a cumulative hour (as $T^c = 60$ minutes) has been spent in the normal zone.

\section{Conclusion and Future Work} \label{sect:conclusions}
In this paper, we presented the CMP formulation of SCED that explicitly accounts for the number of transmission lines that are operating in the emergency zones during a contingency event. The objective function in this new formulation included the total operating cost as well as differential penalties on the number of lines operating in different emergency zones. Constraints ensured that the duration of operation in emergency zones was within the acceptable reliability standards set by the system operators. The CMP is a non-convex, non-smooth optimization problem. We presented two alternative approaches to a solve the CMP. The first approach resulted in a difference-of-convex approximation of the CMP and we used the DCA to obtain its solution. The second approach resulted in a MIP formulation. The numerical experiments illustrated the advantages of using the CMP models in reducing the total operating cost.

The model and the rolling-horizon setup presented in \S\ref{sect:rollingHorizon} accommodate any choice of model resolution and horizon length $T^\prime$. The CMP model instances used in our experiments used $T^\prime=0$. While this reflects the current practice at the ISOs (see \cite{CAISO2017b}), determining an appropriate choice of $T^\prime$ is critical to harness the full potential of lookahead models. A thorough investigation of the SCED model horizon, in general, and the CMP model horizon in particular, is a fruitful research direction.

The current CMP formulation of SCED considers static thermal ratings and therefore, results in a deterministic optimization problem. Several recent studies have shown the advantage of using dynamic thermal rating of transmission lines (e.g., \cite{kazerooni2011dynamic} and \cite{zhan2018stochastic}).  Inclusion of dynamic thermal rating in the CMP model has the potential to further improve system operations. However, this inclusion will result in non-convex, non-smooth stochastic optimization problems. The stochastic difference-of-convex optimization is in its infancy and is the subject of our ongoing work. We will undertake the CMP formulation of SCED with dynamic thermal rating in our future research endeavors.

\bibliographystyle{plain}
\bibliography{EDwTR_Refs.bib}

\end{document}